\documentstyle[preprint,aps]{revtex}
\begin{document}
\draft
\title{Schwinger boson mean field theory of the  Heisenberg Ferrimagnetic
Spin Chain
}

\author{Congjun Wu$^{\dag\ddag}$, Bin Chen$^\ddag$, Xi Dai$^\ddag$, Yue Yu$^\ddag$,
and Zhao-Bin Su$^\ddag$ }
\address{
$^\dag$ Department of Physics, Peking University\\ 
$^\ddag$ Institute of Theoretical Physics, Academia Sinica,
P.O. Box 2735, Beijing 100080, China
}
\maketitle

\begin{abstract}
The Schwinger boson mean field theory is applied to the quantum ferrimagnetic
Heisenberg chain. 
There is a ferrimagnetic long range order  in the ground state.
We observe two branches of the low lying excitation and calculate
the  spin reduction, the gap of the antiferromagnetic branch, and
the spin fluctuation at $T=0K$.
These results agree with the established numerical results quite well.
At finite temperatures, the long range order is destroyed because of 
the disappearance of the Bose condensation. 
The thermodynamic observables, such as the free energy, magnetic susceptibility,
specific  heat, and the spin correlation at $T>0K$, are calculated.
The $T\chi_{uni}$ has a minimum at  intermediate temperatures and the spin
correlation length behaves as $T^{-1}$ at  low temperatures.
These qualitatively agree with the  numerical results and the difference is
small at low temperatures.

\end{abstract}

\vspace*{.8cm}
{\bf Key words}: ferrimagnetic Heisenberg spin chain, Schwinger boson mean
 field theory

{\bf Contact Author}: Congjun Wu

{\bf Mailing Address}: Institute of Theoretical Physics,
 P.O. Box 2735, Beijing 100080,\\
\hspace*{4cm}  P. R. of China

{\bf E-mail}: wucj@itp.ac.cn

\newpage

\centerline{\bf I. INTRODUCTION}
\vspace{.5cm}

A variety of exotic physical phenomena in the low dimensional magnetic
systems have been attracting much interest in recent years.
In these systems, the physical pictures obtained from the classical approach
are often greatly modified or even in contradiction with
the result of the strong quantum fluctuation and topological effect.
Haldane\cite{Hal} conjectured that the one-dimensional integer-spin chain with the nearest
neighbor coupling has an energy gap 
in the spin excitation spectrum and the spin correlation decays exponentially
with distance,
whereas that of the half-odd-integer spin chain is gapless and the spin correlation 
decays algebraically with distance.

It is very interesting to discuss the physical phenomena for  
the spin chain mixed by different kinds of spins.
Recently, one-dimensional Heisenberg ferrimagnetic spin chain, which is
made of two kinds of spins, $S^A=1/2$ and $S^B=1$, has been considered.
\cite{Bre,Matr,Exc,Thermo,DMRG,Tian}.
This one-dimensional chain can be described by the Hamiltonian:
\begin{equation}
H=\sum_{i=1,\eta}^N {\vec S_i^A} \cdot {\vec S_{i+\eta}^B}~,
\end{equation}
where the antiferromagnetic coupling energy $J$ is set to equal 1.
N is the number of unit cells, and $\eta$ is the index of the nearest neighbors.
Brehemer \cite{Bre} {\it et al }  showed that the absolute
ground state of this model has a ferrimagnetic long-range order and obtained
the low-lying excitation,
by using the spin-wave theory(SWT) and the Quantum Monte Carlo method[QMC],
which is confirmed by 
Kolezhuk\cite{Matr} {\it et al} with the matrix product approach.
It is  also consistent with Tian's \cite{Tian} 
rigorous theorem that  the absolute ground state for one-dimensional 
antiferromagnetic Heisenberg model  with unequal spins has 
both antiferromagnetic and ferromagnetic long-range orders.
Furthermore, Yamamoto {\it et al}\cite{DMRG} used the modified 
spin wave theory(MSWT), the density matrix renormalization group method(DMRG)
and QMC to calculate the thermodynamic observables.

In this paper we study the ferrimagnetic spin chain by means of
the Schwinger boson mean field theory(SBMFT). 
The theory has been applied successfully to the infinite Heisenberg chain
\cite{AA},presumably due to the neglect of topological excitations in the
SBMFT.
It can also be extended to the case of the magnetic order range\cite{BC}
by identifying the magnetic order with the Bose condensation of 
the Schwinger bosons.
In the Heisenberg ferrimagnetic spin chain, we find that the SMBFT theory is
suitable to describe both the ground state with the ferrimagnetic long rang 
order and the thermodynamic properties at  finite temperatures.
The mean field theory gives rise to the leading term 
in a systematic 1/N expansion \cite{AA} and
the effects of fluctuation beyond the mean-field theory can also be discussed,
as we will mention later.

In our SBMFT approach, the ground state has a long range ferrimagnetic order
arising from the condensation of the Schwinger bosons at T=0K. 
There are two different kinds of excitation: One is gapless and ferromagnetic
and the other is gapful and antiferromagnetic, which has been pointed out
by Brehmer {\it et al}\cite{Bre}.
The spin reduction, the gap of the antiferromagnetic branch
and the spin correlation at $T=0K$ are calculated and the results are
in good agreement with those of QMC and DMRG\cite{Bre,DMRG}.
When $T>0K$, the two branches of excitation are both gapful,
so that the Bose condensation disappears and there is no real long range order.
This is just what the Wigner-Mermin Theorem  tells us for one dimension.
The gap of the ferromagnetic branch is proportional to $T^2$ and the spin correlation
length is divergent as $1/T$.
The thermodynamic observables, such as the free energy, magnetic susceptibility 
and the specific heat, are calculated.
They agree with the numerical results\cite{Bre,DMRG} qualitatively
and the difference is small at low temperatures.

This paper is organized as follows:
In the second section we construct the mean field theory of the ferrimagnetic
chain.
In the third section we give the ground state properties.
In the fourth section we study the thermodynamic properties. 
Conclusions are made and advantages and limitations compared with 
other approaches are discussed in the final section.

\bigskip
\centerline{\bf II. SCHWINGER BOSON MEAN FIELD THEORY OF }
\centerline {\bf THE FERRIMAGNTIC HEISENBERG SPIN CHAIN}
\vspace{.5cm} 

The spin operator ${\vec S_i^A}$ can be represented by the Schwinger bosons
$a_{i,\uparrow}, a_{i,\downarrow}$ 

\begin{eqnarray}
&&
S^A_{i,+}=a^\dag_{i,\uparrow} a_{i,\downarrow } \qquad
S^A_{i,-}=a^\dag_{i,\downarrow} a_{i,\uparrow }\nonumber  \\
&&
S^A_{i,z}={1\over 2}(a^\dag_{i,\uparrow} a_{i,\uparrow}
-a^\dag_{i,\downarrow} a_{i,\downarrow})~, 
\end{eqnarray}

with
$
S^A_{i}={1\over 2}(a^\dag_{i,\uparrow} a_{i,\uparrow}
+a^\dag_{i,\downarrow} a_{i,\downarrow})~ 
$ on each site of kind A.
And ${\vec S_j^B}$ can be represented in a similar way.

The Hamiltonian(1) is rewritten in this representation
\begin{mathletters}
\begin{eqnarray}
&&
 H=-{1\over 2}\sum_{i=1,\eta}^N 
 (a^\dag_{i,\uparrow}b^\dag_{i+\eta,\downarrow}-a^\dag_{i,\downarrow}b^\dag_{i+\eta,\uparrow})
 (a_{i,\uparrow}b_{i+\eta,\downarrow}-a_{i,\downarrow}b_{i+\eta,\uparrow})
+\sum_{i,\eta}S^A_iS^B_{i+\eta}  \nonumber \\
&&
\qquad
=-2\sum_{i=1,\eta}^N D^\dag_{i,i+\eta}D_{i,i+\eta}+\sum_{i,\eta}S^A_iS^B_{i+\eta}\\
&&
D_{i,i+\eta}={1\over 2}
(a_{i,\uparrow}b_{i+\eta,\downarrow}-a_{i,\downarrow}b_{i+\eta,\uparrow})~.
\end{eqnarray}
\end{mathletters}
Considering the constraint $\sum_{\sigma}a^\dag_{i,\sigma}a_{i,\sigma}
=2S^A$ or $\sum_{\sigma}b^\dag_{j,\sigma}b_{j,\sigma}=2S^B$ on each site,
we may introduce two kinds of Lagrangian multipliers $\lambda^A_i$ and
$\lambda^B_j$ to impose the constraint.
At the mean field lever, we can take the average value of the bond operator 
$<D_{i,i+\eta}>=D$ to be uniform and static.
And so are $<\lambda^A_i>=\lambda^A$ and $<\lambda^B_j>=\lambda^B$.

The mean field Hamiltonian reads
\begin{mathletters}
\begin{eqnarray}
&&
H_{MF}=-\sum_{i=1,\eta}^N 
\{D^*(a_{i,\uparrow}b_{i+\eta,\downarrow}-a_{i,\downarrow}b_{i+\eta,\uparrow})
+(a^\dag_{i,\uparrow}b^\dag_{i+\eta,\downarrow}-a^\dag_{i,\downarrow}b^\dag_{i+\eta,\uparrow})D\}
 \\
&&
\qquad\qquad
+\lambda^A\sum_{i=1,\eta}^N (a^\dag_{i,\sigma}a_{i,\sigma}-2S^A)
+\lambda^B\sum_{j=1,\eta}^N (b^\dag_{j,\sigma}b_{j,\sigma}-2S^B)
+2zND^*D +zNS^AS^B ,
\end{eqnarray}
in momentum space, which is transformed into
\begin{eqnarray}
&&
H_{MF}=\sum_{k,\sigma}\{ \lambda^A a^\dag_{k,\sigma}a_{k,\sigma}
+\lambda^B b^\dag_{k,\sigma}b_{k,\sigma}\}
-\sum_{k,\sigma}\{D^*z\gamma^*_k(a_{k,\uparrow}b_{-k,\downarrow}-a_{k,\downarrow}b_{-k,\uparrow})
\nonumber \\
&&
\qquad\qquad
+Dz\gamma_k(a^\dag_{k,\uparrow}b^\dag_{-k,\downarrow}-a^\dag_{k,\downarrow}b^\dag_{-k,\uparrow})\}
 +2zND^*D-2N(S^A\lambda^A+S^B\lambda^B) +zNS^AS^B   .
\end{eqnarray}
\end{mathletters}
Here z is the number of the nearest neighbor sites 
and equals 2 in the one-dimensional chain,
and $\gamma_k={1\over z}\sum_\eta  e^{ik\eta} =\cos k~.$
The sum of k is restricted in the reduced first Brillouin zone,
 which extends from $-{\pi \over 2}$
to ${\pi \over 2}$.

Using the Bogoliubov transformation 
\begin{equation}
 \left( \begin {array} {c}
           a_{k,\uparrow}\\b^\dag_{-k,\downarrow}
         \end{array} \right) 
= \left( \begin {array}{cc}
         \cosh\theta_k & \sinh\theta_k\\
         \sinh\theta_k & \cosh\theta_k
        \end{array} \right) 
  \left( \begin {array} {c}
         \alpha_{k,\uparrow}\\ \beta^\dag_{-k,\downarrow}
        \end{array} \right) 
~        
  \left( \begin {array} {c}
           b_{k,\uparrow}\\a^\dag_{-k,\downarrow}
         \end{array} \right) 
= \left( \begin {array}{cc}
         \cosh\theta_k & -\sinh\theta_k\\
         -\sinh\theta_k & \cosh\theta_k
        \end{array} \right) 
  \left( \begin {array} {c}
         \beta_{k,\uparrow}\\ \alpha^\dag_{-k,\downarrow}
        \end{array} \right) ,
\end{equation}
with $\theta$ given by
\begin{mathletters}
\begin{equation}
\tanh2{\theta}={|zD\gamma_k| \over (\lambda^A+\lambda^B)/2}~,   
\end{equation}
we obtain the energy spectrum 
\begin{eqnarray}
&&
E^\alpha_{k,\sigma}={\lambda^A-\lambda^B \over 2} +
\sqrt{ ({\lambda^A+\lambda^B \over 2})^2-|zD\gamma_k|^2}  \\
&&
E^\beta_{k,\sigma}=-{\lambda^A-\lambda^B \over 2} +
\sqrt{ ({\lambda^A+\lambda^B \over 2})^2-|zD\gamma_k|^2}~.
\end{eqnarray}
\end{mathletters}
The Hamiltonian is diagonalized and it is easy to write down the free energy,
\begin{mathletters}
\begin{eqnarray}  
&& 
H_{MF}=\sum_{k,\sigma} \{ E^\alpha_k(\alpha^\dag_{k,\sigma}\alpha_{k,\sigma}+1/2)
 +E^\beta_k(\beta^\dag_{k,\sigma}\beta_{k,\sigma}+1/2)\} +2zND^*D \nonumber \\
&& 
\qquad \qquad
-2N(S^A+1/2)\lambda^A-2N(S^B+1/2)\lambda^B + zNS^AS^B\\
&& {F^{MF}\over 2N}={2\over \beta} \int_{-\pi\over 2} ^{\pi \over 2}{ dk\over 2\pi}
\{ \ln(2\sinh({\beta\over 2}E^\alpha_k)+ \ln (2\sinh({\beta\over 2}E^\beta_k) \}
+zD^*D  \nonumber  \\
&&
\qquad \qquad 
-(S^A+1/2)\lambda^A-(S^B+1/2)\lambda^B+ {z\over2}S^AS^B ~ .
\end{eqnarray}
\end{mathletters}
The mean field self-consistent equations can be obtained by minimizing the free
energy,  i.e.
 $\delta F / \delta \lambda^A=0$,
$\delta F / \delta \lambda^B=0$ and $\delta F / \delta D^*=0$.
After a simple algebra, the equations are rearranged as:
\begin{mathletters}
\begin{eqnarray}
&& S^B-S^A=\int_{-\pi\over 2}^{\pi\over 2}{dk\over 2\pi}
\{ \coth{\beta\over 2}E^\beta_k-\coth{\beta\over 2}E^\alpha_k \} \\
&& S^B+S^A+1=\int_{-\pi\over 2}^{\pi\over 2}{dk\over 2\pi}
{ {\lambda^A+\lambda^B \over 2} \over \sqrt{ ({\lambda^A+\lambda^B \over 2})^2
-|zD\gamma_k|^2} }
\{ \coth{\beta\over 2}E^\beta_k+\coth{\beta\over 2}E^\alpha_k \}  \\
&& {2\over z}=\int_{-\pi\over 2}^{\pi\over 2}{dk\over 2\pi}
{ |\gamma_k|^2 \over \sqrt{ ({\lambda^A+\lambda^B \over 2})^2 -|zD\gamma_k|^2} }
\{ \coth{\beta\over 2}E^\beta_k+\coth{\beta\over 2}E^\alpha_k \}
\end{eqnarray}
\end{mathletters}

Rescale the parameters $(\lambda^A,\lambda^B,D,\beta)
\longrightarrow (\Lambda_1,\Lambda_2,\eta,\kappa)$ \cite{AA}
\begin{eqnarray}
&&{\lambda^A+\lambda^B \over 2}={1\over 2}z\Lambda_1,~D={1\over 2}\Lambda_1\eta
\nonumber \\
&&{\lambda^A-\lambda^B \over 2}={1\over 2}z\Lambda_2,~\beta={4\kappa\over z}~.
\end{eqnarray}
Then the angle of the Bogoliubov transformation is expressed in a compact form
\begin{eqnarray}
&&
\cosh 2\theta_k={1\over \sqrt{1-\eta^2 \gamma_k^2} }
~\qquad
\sinh 2\theta_k={|\eta \gamma_k|\over \sqrt{1-\eta^2 \gamma_k^2} },
\end{eqnarray}
and the self-consistent equations read
\begin{mathletters}
\begin{eqnarray}
& &S^B-S^A=\int_0^{\pi\over 2}{dk\over \pi}\{
\coth[\kappa(\Lambda_1\sqrt{1-\eta^2\gamma_k^2}-\Lambda_2) ]-
\coth[\kappa(\Lambda_1\sqrt{1-\eta^2\gamma_k^2}+\Lambda_2) ] \}\\
& &S^B+S^A+1=\int_0^{\pi\over 2}{dk\over \pi}\{
{ \coth[\kappa(\Lambda_1\sqrt{1-\eta^2\gamma_k^2}-\Lambda_2) ]+
  \coth[\kappa(\Lambda_1\sqrt{1-\eta^2\gamma_k^2}+\Lambda_2) ]  
\over \sqrt{1-\eta^2\gamma_k^2} } \}\\
& &S^B+S^A+1-\Lambda_1\eta^2=\int_0^{\pi\over 2}{dk\over \pi}\{
\coth[\kappa(\Lambda_1\sqrt{1-\eta^2\gamma_k^2}-\Lambda_2) ]+\nonumber\\
& &     \qquad \qquad \qquad \qquad \ \qquad                                                   \
 \coth[\kappa(\Lambda_1\sqrt{1-\eta^2\gamma_k^2}+\Lambda_2) ] \}~
\sqrt{1-\eta^2\gamma_k^2}~.
\end{eqnarray}
\end{mathletters}

\bigskip
\centerline{\bf III.THE PROPERTIES OF THE GROUND STATE     }
\vspace{.5cm} 

Notice that the Bogoliubov particles in the
$\beta$ branch have to condense to the ground state at $T=0K$
as long as $S^A\not=S^B$,
otherwise the equation (11.a) cannot be satisfied.
The excitation energy $E_\beta$ has its minimal value $E_\beta=0$ at $k=0$ 
while the $\alpha-$branch has a gap of 2$\Lambda_2$ at $T=0K$. 
Sarker {\it et al}\cite{BC} showed that the long range
order is related to the condensation of the Schwinger bosons in both the 
ferromagnetic and antiferromagnetic Heisenberg models. 
We, now, arrive the same conclusion for the ferrimagnetism model
at one dimension.
This can be contrasted to the antiferromagnetic case,
where there is no long range order even at $T=0K$ at one dimension.

Suppose that there is an infinitesimal external stagger magnetic
field which is upward at the B site and downward at the A site.
Then the $\beta_\uparrow$-branch has the lowest energy and the bosons condense
at the state of $\beta_{\uparrow,k=0}$.
Because of the bose condensation, the self-consistant equations at $T=0K$
are modified as follows:
\begin{mathletters}
\begin{eqnarray}
&&
 S^B-S^A={1\over 4N}\coth[\kappa (\Lambda_1\sqrt{ 1-\eta^2}-\Lambda_2)]|_{
\kappa\rightarrow+\infty} \\
&&
\Lambda_1\sqrt{ 1-\eta^2}=\Lambda_2 \\
&&
S^B+S^A+1={2\over \pi}K(\eta)+{(S^B-S^A) \over \sqrt{ 1-\eta^2} }\\
&&
S^B+S^A+1-\Lambda_1 \eta^2={2\over \pi}E(\eta)+(S^B-S^A) \sqrt{ 1-\eta^2}
\end{eqnarray}
\end{mathletters}
Here $E(\eta)$ and $K(\eta)$ are the first and second type complete elliptic integrals.
The parameters have been determined numerically for the case in $S^A=1/2$ and $S^B=1$ with
$\eta=0.8868, \Lambda_1=1.9238,$ and $\Lambda_2=0.8890$.

The average value of the spin on the site B ,
\begin{eqnarray}
&&
<S^B_Z>={1\over2}<b^\dag_{j\uparrow}b_{j\uparrow}-b^\dag_{j\downarrow}b_{j\downarrow}>
={1\over 2N}\sum_k<b^\dag_{k\uparrow}b_{k\uparrow}-b^\dag_{k\downarrow}b_{k\downarrow}>~,
\nonumber 
\end{eqnarray}
can be calculated at $T=0K$,
by using the quasi-particle operator 
\begin{mathletters}
\begin{eqnarray}
&&
<S^B_Z>={1\over 2N}\cosh^2\theta_k n_{\beta k\uparrow}|_{k=0, T\rightarrow 0K}
={1\over 2}(1+{1\over \sqrt{1-\eta^2}})(S^B-S^A)=0.791.
\end{eqnarray}
Similarly, the average value of the spin on the site A is
\begin{eqnarray}
&&
<S^A_Z>=-{1\over 2N}\sinh^2\theta_k n_{\beta k\uparrow}|_{k=0, T\rightarrow 0K}
={1\over 2}(1-{1\over \sqrt{1-\eta^2}})(S^B-S^A)=-0.291.
\end{eqnarray}
The spin reduction $\tau=S^B-<S^B_z>$ on the site B
or $\tau=S^A+<S^A_z>$ on the site A is given by
\begin{eqnarray}
&&
\tau^A=\tau^B={(S^B+S^A)\over 2}-{1\over \sqrt{1-\eta^2} }{(S^B-S^A)\over 2}
=0.209~.
\end{eqnarray}
\end{mathletters}

From equation(11a), we can see that the number of the condensed bosons on the state
$\beta_{\uparrow,k=0}$ is $2N(S^B-S^A)$, which is just the number of the Schwinger
Bosons on the site B  subtracted by that on the site A. 
As long as $S^A\not=S^B$, there is
a ferrimagnetic long range order, which agrees with Tian's rigorous proof\cite{Tian}.
We quote several known results for the value of $\tau$ to  
show the satisfactory of our calculation.
The QMC \cite{Bre}  gives $\tau=0.207\pm0.002$,and $\tau=0.221$ in the matrix product states
approach \cite{Matr}.
The naive SWT overestimates the spin reduction and results in $\tau=0.3$ \cite{Bre}.

The gap of the antiferromagnetic branch in our mean field theory is 
$\Delta_{anti}=2\Lambda_2=1.778$, which is very close to that in the exact-diagonalization 
\cite{Exc} $\Delta_{anti}=1.759$ and in the QMC \cite{Bre},
$\Delta_{anti}=1.767$.
The naive SWT\cite{Bre} and the MSWT\cite{DMRG} give the gap  $\Delta_{anti}=1$ and
$\Delta_{anti}=1.676$, respectively.

We calculate the ground state energy of one unit cell, which yields the zero temperature 
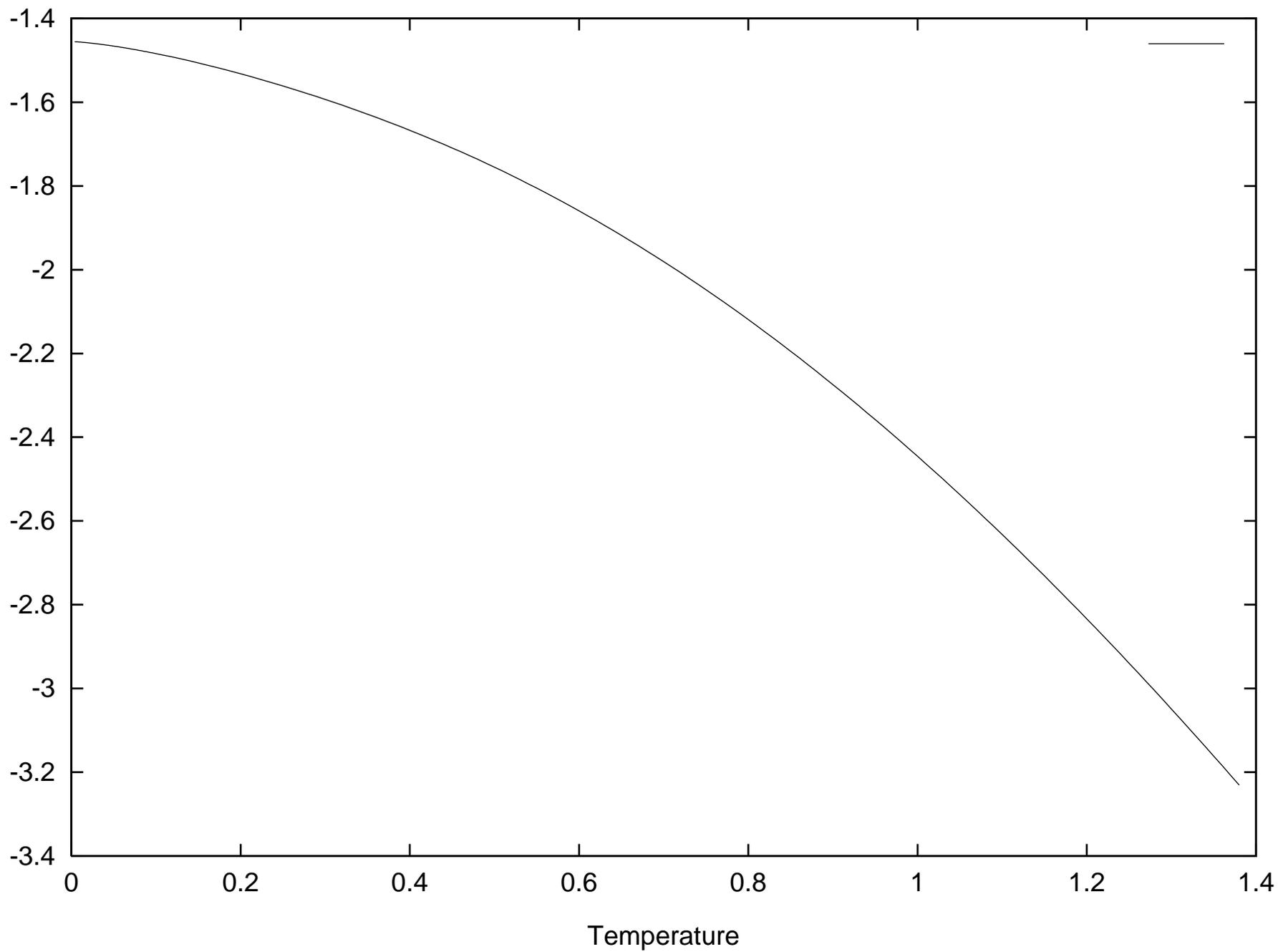
free energy per unit cell, 
\begin{eqnarray}
&&
{F_{MF}\over N}= \int_{-\pi\over 2} ^{\pi \over 2}{ dk\over 2\pi}
\{4\Lambda_1\sqrt{1-\eta^2\gamma_k^2}\}+{z\over2}\Lambda_1^2\eta^2
-2\Lambda_1(S^A+S^B+1)-2\Lambda_2(S^A-S^B)+zS^AS^B  \nonumber \\
&&
\qquad=-1.904
\end {eqnarray}
This result is much lower than those of the QMC \cite{Bre}
and the MSWT \cite{DMRG}, which are  $-1.437$ and $-1.454$, respectively.
In fact, this is an artificial result caused by the mean field theory,
in which  we assume that  the constraint and
the bond operator are uniform and static.
We have overcounted the degrees of freedom of the Schwinger bosons by a factor of 2,
as argued by Arovas and Auerbach\cite{AA,AAl}.
To count the degrees of freedom correctly, we have to divide the 
part of fluctuation per unit cell ${F_{MF}/ N}+2S^AS^B$ by 2,
and add back the classic ground energy per cell $-2S^AS^B$.
Then we  have  the modified result $-1.455$.
The $1/N$ expansion can give the above argument a strong basis.

The spin correlation length $\xi$ of the ground state is also of  interest.
Because of the appearance of the long range order, the transverse and longitudinal
fluctuations are  anisotrope.
In our SBMFT, the longitudinal correlation between  
the site A and the site B can be calculated as
\begin{mathletters}
\begin{eqnarray}
&&
<S^A_{i,z}S^B_{j,z}>-<S^A_{i,z}><S^B_{j,z}>\nonumber \\
&&
=-<(S^A_{i,z}-a^\dag_{i,\uparrow}a_{i,\uparrow})
(S^B_{j,z}-b^\dag_{j,\downarrow}b_{j,\downarrow})>+
<(S^A_{i,z}-a^\dag_{i,\uparrow}a_{i,\uparrow})>
<(S^B_{j,z}-b^\dag_{j,\downarrow}b_{j,\downarrow}> \nonumber \\
&&
=-<a^\dag_{i,\uparrow}a_{i,\uparrow} b^\dag_{j,\downarrow}b_{j,\downarrow}>
+<a^\dag_{i,\uparrow}a_{i,\uparrow}> <b^\dag_{j,\downarrow}b_{j,\downarrow}>
\nonumber \\
&&
=-|g(R_{ij})|^2\\
&&
g(R_{ij})=\int_{-\pi\over 2}^{\pi\over 2}{dk \over 2\pi}\sinh2\theta_k e^{-ikR_{ij}}
~.
\end{eqnarray}
Similarly, the longitudinal fluctuations between A site and A site and 
between B site and B site can be given  by
\begin{eqnarray}
&&
<S^A_{i,z}S^A_{j,z}>-<S^A_{i,z}><S^A_{j,z}>=|f(R_{ij})|^2\\
&&
<S^B_{i,z}S^B_{j,z}>-<S^B_{i,z}><S^B_{j,z}>=|f(R_{ij})|^2\\
&&
f(R_{ij})=\int_{-\pi\over 2}^{\pi\over 2}{dk \over 2\pi}\cosh2\theta_k e^{-ikR_{ij}}
~.
\end{eqnarray}
\end{mathletters}
From equations(15.b,15.e) we get the correlation length $\xi=\eta / [8(1-\eta^2)]^{1/2}=0.6785$.
Although the correlation functions are not in a good exponential form because $\eta$ is not 
close to 1, we can still see that the correlation decays very rapidly.

The there kinds of transverse correlation are given by
\begin{mathletters}
\begin{eqnarray}
&&
<S^A_{i,+}S^B_{j,-}>=<a^\dag_{i\uparrow}a_{i\downarrow}
b^\dag_{j\downarrow}b_{j\uparrow}>
=-|g(R_{ij})|^2-(S^B-S^A) g(R_{ij}) \sinh2\theta_k|_{k=0}~,\\
&&
<S^A_{i,+}S^A_{j,-}>=<a^\dag_{i\uparrow}a_{i\downarrow}
a^\dag_{j\downarrow}a_{j\uparrow}>
=|f(R_{ij})|^2+2(S^B-S^A)f(R_{ij})\sinh^2\theta_k|_{k=0}~,\\
&&
<S^B_{i,+}S^B_{j,-}>=<b^\dag_{i\uparrow}b_{i\downarrow}
b^\dag_{j\downarrow}b_{j\uparrow}>
=|f(R_{ij})|^2+2(S^B-S^A)f(R_{ij})\cosh^2\theta_k|_{k=0}~.
\end{eqnarray}
\end{mathletters}
We note that the transverse correlation length is 2 times of the longitudinal one.
The SWT calculation gives that the longitudinal correlation length 
$\xi =1/(2\ln2)=0.7213$\cite{Bre}, while the numerical methods 
cannot give the accurate correlation length, because the fluctuation
decays so rapidly.

\bigskip
\centerline{\bf IV. THE THERMODYNAMIC PROPERTIES AT FINITE TEMPERATURES}
\vspace{.5cm} 

We investigate the low temperature asymptotic expansion of the self-consistent equations(11)
and verify that it changes continuously into the equations (12) in $T=0$.
The equation (11.a) has a solution for $T\not=0K$. 
So there is no boson condensation.
The equation (11.a) gives the gap of the ferromagnetic branch as
$$
\Delta_{ferro}={\sqrt{1-\eta^2}\over \eta^2}{2T^2\over (S^B-S^A)^2\Lambda_1}
\approx 2.4436T^2.
$$

We  solved the self-consistant equations (11) numerically and find that
the values of the variation parameters are continuously evolved to the values
at T=0K.
On the other hand, we see that the mean field theory fails and the bond operator $D$ is zero 
when T is higher than a specific temperature, say approximately 1.38 in our case.
The failure of the mean field indicates that the system has entered a local
moment phase in which there is no correlation of the spin fluctuation, as
pointed out by Aromas {it et al} \cite{AA}.
With temperature increasing, we find that the gap of the antiferromagnetic
branch $\Delta_{anti}=\Lambda_2+\Lambda_1\sqrt{1-\eta^2}$  varies.
In the temperature region where the SBMFT is valid, this gap firstly decreases
and then increases.
It reaches its minimum which is about 1.30 when $T\approx 0.7$.
The gaps of the ferromagnetic and antiferromagnetic branches are plotted
in Figure.~1.
And the free energy {\it versa} T is calculated and plotted in Figure.~2 
with the argument of dividing the part of fluctuation by 2 ~\cite{AA,AAl}.

The spin correlation at $T\rightarrow 0K$ are calculated:
\begin{mathletters}
\begin{eqnarray}
&&
<{\vec S^A_i}\cdot{\vec S^B_j}> =-{3\over 2}|G(R_{ij})|^2 \nonumber \\
&&
G(R_{ij})={1\over 2} \int_{-\pi\over 2}^{\pi\over 2}{dk \over 2\pi}
\sinh 2\theta_k \{\coth{\beta\over 2}E^\alpha_k+\coth{\beta\over 2}E^\beta_k\}
e^{-ikR_{ij}}~,\\
&&
<{\vec S^A_i}\cdot{\vec S^A_j}> ={3\over 2}|F_1(R_{ij})|^2  \nonumber \\
&&
F_1(R_{ij})= \int_{-\pi\over 2}^{\pi\over 2}{dk \over 2\pi}
\{ \cosh^2 \theta_k \coth{\beta\over 2}E^\alpha_k+
\sinh^2 \theta_k \coth{\beta\over 2}E^\beta_k \} e^{-ikR_{ij}}~,\\
&&
<{\vec S^B_i}\cdot{\vec S^B_j}> ={3\over 2}|F_2(R_{ij})|^2  \nonumber \\
&&
F_2(R_{ij})= \int_{-\pi\over 2}^{\pi\over 2}{dk \over 2\pi}
\{ \sinh^2 \theta_k \coth{\beta\over 2}E^\alpha_k+
\cosh^2 \theta_k \coth{\beta\over 2}E^\beta_k \} e^{-ikR_{ij}}~.
\end{eqnarray}
\end{mathletters}
Because $\Delta_{ferro}$ behaves as $T^2$, the correlation length
behaves as $T^{-1}$  i.e. $\xi=\eta^2 \Lambda_1(S^B-S^A)/ (4\sqrt{1-\eta^2})
T^{-1}$.

The dynamic magnetic susceptibility is calculated by using linear response
theory,
\begin{equation}
 \left( \begin {array} {c}
          \Delta S^A_z(q,\omega)\\ \Delta S^A_z(q,\omega)
         \end{array} \right) 
~=-g
\left( \begin {array} {cc}
        <<S^A_zS^A_z>>_{q\omega}& <<S^A_zS^B_z>>_{q\omega}\\
        <<S^B_zS^A_z>>_{q\omega}& <<S^B_zS^B_z>>_{q\omega}
         \end{array} \right) 
~
\left( \begin {array} {c}
         h_A(q,\omega)\\h_B(q,\omega)
         \end{array} \right)        
\end{equation}~,
where $<<S^A_zS^A_z>> {\it etc. }$ is the retarded Green function, and $h_A,h_B$ 
are the small external magnetic field, $g$ is the Laude factor.
In the Matsubra representation, the dynamic magnetic susceptibilities are given by
\begin{mathletters}
\begin{eqnarray}
&&
\chi^{zz}_{AA}(q,i\omega_n)=-g^2<<S^A_zS^A_z>>_{q\omega} 
={g^2\over 2N}\sum_k \{ { n_B(E^\alpha_k)-n_B(E_{k+q}^\alpha) 
\over i\omega_n+E_{k+q}^\alpha-E_k^\alpha }
\cosh^2\theta_{k+q}\cosh^2\theta_k \nonumber \\
&&
+{ n_B(E_{k+q}^\beta)-n_B(E_q^\beta) \over i\omega_n-E_{k+q}^\beta+E_k^\beta }
\sinh^2\theta_{k+q}\sinh^2\theta_k
+{1+n_B(E_k^\beta)+n_B(E_{k+q}^\alpha) \over i\omega_n+E_{k+q}^\alpha+E_k^\beta }
\cosh^2\theta_{k+q}\sinh^2\theta_{k}  \nonumber \\
&&
-{1+n_B(E_k^\alpha)+n_B(E_{k+q}^\beta) \over i\omega_n-E_{k+q}^\beta-E_k^\alpha }
\cosh^2\theta_{k}\sinh^2\theta_{k+q} \} ~,\\
&&
\chi^{zz}_{AB}(q,i\omega_n)=-g^2<<S^A_zS^B_z>>_{q\omega} 
={g^2\over 2N}\sum_k \sinh\theta_k\sinh\theta_{k+q}\cosh\theta_k\cosh\theta_{k+q}
\nonumber\{  \\
&&
\qquad \qquad\qquad \qquad
{ 1+n_B(E_{k+q}^\beta)+n_B(E_k^\alpha) \over i\omega_n-E_{k+q}^\beta-E_k^\alpha }
+{n_B(E^\alpha_{k+q})-n_B(E^\alpha_k) \over i\omega_n+E^\alpha_{k+q}-E^\alpha_k }
\nonumber \\
&&
\qquad \qquad\qquad \qquad
+{n_B(E^\beta_k)-n_B(E^\beta_{k+q}) \over i\omega_n+E^\beta_k-E^\beta_{k+q} }
-{1+n_B(E^\alpha_{k+q})+n_B(E^\beta_k) \over i\omega_n+E^\alpha_{k+q}+E^\beta_k }
\} ~.
\end{eqnarray}
\end{mathletters}
$\chi^{zz}_{BB}(q,i\omega_n)$ and $\chi^{zz}_{BA}(q,i\omega_n)$ can be obtained
by the exchange $(\alpha\leftrightarrow\beta)$ in the equation (19.a) and(19.b)
respectively.

The mean field static uniform and staggered magnetic susceptibilities per unit cell are
\begin{mathletters}
\begin{eqnarray}
&&
{\chi_{uni}^{MF}\over N }=g^2\beta \int_{-\pi\over 2}^{\pi\over 2}{dk \over 2\pi}
\{ n^\alpha_k(n^\alpha_k+1)+n^\beta_k(n^\beta_k+1)\}  \\
&&
{\chi_{stag}^{MF}\over N }=g^2\beta \int_{-\pi\over 2}^{\pi\over 2}{dk \over 2\pi}
\{ [n^\alpha_k(n^\alpha_k+1)+n^\beta_k(n^\beta_k+1)] \cosh^2 2\theta_k
+2\sinh^2 2\theta_k{1+n^\alpha_k+n^\beta_k \over \beta(E^\alpha_k+E^\beta_k) }
\}~.
\end{eqnarray}
If $S^A=S^B$, the above equations are reduced to the familiar forms of the
antiferromagnetic case \cite{AA}, which are:
\begin{eqnarray}
&&
{\chi_{uni}^{MF}\over 2N }=g^2\beta \int_{-\pi\over 2}^{\pi\over 2}{dk \over 2\pi}
n_k(n_k+1)~,\\
&&
{\chi_{stagger}^{MF}\over 2N }=g^2\beta \int_{-\pi\over 2}^{\pi\over 2}{dk \over 2\pi}
\{ n_k(n_k+1)\cosh^2 2\theta_k+{n^\beta_k+1/2 \over \beta E^\beta_k} \sinh^2 2\theta_k
\}~.
\end{eqnarray}
\end{mathletters}
For low temperatures, both  $\chi_{uni}^{MF}$ and $\chi_{stag}^{MF}$ 
are proportional to $T^{-2}$, {\it e.g.} $\chi_{uni}^{MF}=g^2
\Lambda_1\eta^2(S^B-S^A)^3/( 2\sqrt{ (1-\eta^2) } )T^{-2}$, and so on.
We plot ${T\chi_{uni} / Ng^2}$ versa $T$ in Figure.~3 and ${T\chi_{stag} / Ng^2}$ versa $T$ 
in Figure.~4.
There we have multiply the mean field susceptibilities with the factor
${2\over 3}$ due to the same argument as Arovas and Auerbach\cite{AA,AAl}.
$T\chi_{uni}/Ng^2$ reaches a minimum of 0.4 at the intermediate 
temperature region around $T\approx 0.5$.
This is due to the contribution from the gapful antiferromagntic branch.
The low temperature behavior of $T\chi_{uni}/Ng^2$ when $T<0.4-0.5$, 
the location and the value
of the minimum are found a good agreement with
with the QMC and DMRG calculations, even better than the 
MSWT calculation with improved dispersion relations \cite{DMRG}.
After the point of minimum, the SBMFT result increases too rapidly, 
showing a discrepancy with numerical calculations.
$T\chi_{stag}/Ng^2$ is dropped rapidly and monotonously with the temperature 
increasing.

The specific heat ${C / N}$  ${\it versa}$  T is calculated by the numerical 
differentiation of the internal energy with T, and is plotted in Figure.~5. 
We also find  the low temperature behavior of $C \propto T^{1/2}$,
which agrees with the  QMC and DMRG calculation \cite{Thermo,DMRG} well
when $T<0.4K$.
Again the SBMFT result increases rapidly,
failing to see the Shocttky-like peak\cite{DMRG} at
intermediate temperatures.

In short, the SBMFT result describes the thermodynamic properties  
well at low temperatures($T<0.4-0.5$).
The disagreement in intermediate and high temperatures
is also  owing to  the static and uniform 
constraint.

\bigskip
\centerline{\bf V. CONCLUSIONS}
\vspace{.5cm} 

Using the Schwinger boson mean field theory, 
we have investigated both
the ground state  and the thermodynamic properties of the Heisenberg 
ferrimagntic spin chain.
The long range ferrimagnetic order of the ground state is caused by 
the condensation of the Schwinger bosons.
The spin fluctuations to the ground state are anisotropic 
and decays very rapidly.
The excitation spectrum has both the low energy ferromagnetic branch which 
is gapless at $T=0K$ and the high energy antiferromagnetic branch with
a gap is 1.778 at $T=0K$.
With the temperature increasing, the branch of the ferromagnetic become gapful.
$\Delta_{ferro}$ behaves as $T^2$ and $\Delta_{anti}$ varies.
At low temperatures, the ferrimagnetic spin chain exhibits the feature of the
ferromagnetism.
The static magnetic susceptibility, the spin correlation length and 
the specific heat behave as $T^{-2}$, $T^{-1}$ and $T^{1/2}$, respectively.
At intermediate temperatures, the antiferromagnetic branch begins to play 
an important role\cite{DMRG}.
The $T\chi_{uni}$ has a minimum at $T\approx 0.5$.

Compared with other approaches, the SBMFT is a simple mean field theory
but can give many good results at both zero and finite temperatures.
The spin reduction and gap of the antiferromagnetic branch
$\Delta_{anti}$ at $T=0K$  differ from the
numerical calculations less than $1\%$.
The thermodynamic properties, such as the  static uniform magnetic 
susceptibility $\chi_{uni}$ and the specific heat $C$ 
calculated by the SBMFT, agree with numerical results well when $T<0.4$.
The spin correlation which is anisotropic at the ground state and 
isotropic at finite
temperatures can be calculated easily.
These results improve those of the SWT\cite{Bre} largely and 
are consistent with those of the more complicated MSWT
and numerical methods\cite{Thermo,DMRG}.

The SBMFT is not successful at intermediate and high temperatures.
The behavior of  $\chi_{uni}$ has quantitative discrepancy with
numerical results, and that of $C$ does not agree with 
numerical calculations qualitatively when $T>0.4$.
Both of them increase too rapidly with temperatures increasing.
When $T>1.38$, the order parameter drops to zero and the SBMFT theory fails.
It can not describe the system in the whole temperature range as
the numerical methods and the MSWT do.

\bigskip
\centerline{\bf ACKNOWLEDGMENTS}
\vspace{.5cm} 
We thank Professor Tian Guangshan for his
stimulating remarks.

\newpage
\centerline{\bf Figure Caption}
\vspace{.5cm} 

Fig.1 The gaps of the ferromagnetic and antiferromagnetic branches

Fig.2 The free energy $F$ {\it versa} T.

Fig.3 The $T\chi_{uni} /Ng^2$ {\it versa} T.

Fig.4 The $T\chi_{stag} /Ng^2$ {\it versa} T.

Fig.5 The specific heat {\it versa} T.

\end{document}